\documentstyle[epsfig]{jaa}
\input{def.lst}
\volume{20}
\pubyear{1999}
\pagerange{\pageref{firstpage}--\pageref{lastpage}}
\begin{document}
\setcounter{page}{0}
\label{firstpage}
\title[The interstellar clouds of Adams and Blaauw - I]{The interstellar clouds of 
Adams and Blaauw revisited: an HI absorption study - I}

\author[Rajagopal, Srinivasan \& Dwarakanath]{Jayadev Rajagopal, G. Srinivasan and K.S. Dwarakanath\\ 
Raman Research Institute,\\
               Bangalore 560 080, India.\\
       email: jayadev@stsci.edu, srini@rri.ernet.in, dwaraka@rri.ernet.in}

\date{Received ; Accepted }
   \maketitle
\begin{abstract}
This investigation is aimed at clarifying the nature of the interstellar 
gas seen in absorption against bright O and B stars. Towards this end we
have obtained for the first time HI absorption spectra towards radio sources
very close to the lines of sight towards 25 bright stars previously studied.
In this paper we describe the selection criteria, the details regarding our
observations, and finally present the absorption spectra. In the accompanying
paper we analyse the results and draw conclusions.
\end{abstract}

\section{Introduction}
We have carried out an absorption study in the 21 cm line of atomic
hydrogen in 25 directions
in the Galaxy. These directions have been selected from previous
optical absorption studies in the lines of singly ionized calcium (CaII)
and neutral sodium (NaI).
In this paper, we describe the observations and present the data obtained
by us.
A discussion of the results and the conclusions drawn from the study 
are in the accompanying paper (Rajagopal, Srinivasan and 
Dwarakanath 1998; paper II).

Our observations were primarily intended to study the velocities of the 
HI absorption features, their linewidths, and in combination with
existing HI emission measurements obtain the spin temperature of the absorbing gas. 
Our study was motivated by the following questions:
\begin{itemize}
\item[(i)] How are interstellar clouds seen in optical absorption related
to those seen in HI absorption and emission ?
\item[(ii)] What is the nature of the relatively fast clouds so commonly
seen in the absorption lines of NaI and CaII ?
\end{itemize}
To clarify these and related issues we briefly summarize the salient
historical background.
Some of the earliest information about the Interstellar Medium (ISM) came
from observations of optical absorption in the H and K ($\lambda\lambda 3933, 3968{\rm\AA}$)
lines of CaII, and the D$_1$ and D$_2$ ($\lambda\lambda 5889, 5895{\rm\AA}$) lines of
NaI towards bright stars. Adams (1949)
made an extensive study of the absorption lines of CaII towards
nearly 300 O and B stars. 
These observations 
were later extended to the D lines of NaI (Routly and Spitzer 1952; Hobbs 1969), 
and to high latitude stars (M\"{u}nch and Zirrin 1961).In the simplest model, these 
lines were attributed 
to interstellar gas in the form of clouds. 
The existence of a {\em hot intercloud medium} was conjectured by Spitzer (1956)
and the theoretical basis for a two-phase model followed ( Field 1965; Field, Goldsmith
and Habing 1969).

Independently, a global picture of the ISM emerged from radio
observations of the 21 cm HI line (Clark, Radhakrishnan and Wilson 1962; Clark 1965;
Radhakrishnan et. al. 1972).
These and later studies have established that an important constituent
of the ISM are cool diffuse clouds in pressure equilibrium with a warmer intercloud medium.
The notion of interstellar ``clouds'' was thus invoked to explain both
optical and radio observations. {\em But one was left speculating as to whether
the two populations were the same.} Surprisingly the answer to the above question still remains incomplete.
Many properties of the two populations seem to differ. In particular the
number of clouds per kiloparsec (Blaauw 1952; Radhakrishnan and
Goss 1972; Hobbs 1974; Radhakrishnan and Srinivasan 1980), and the velocity distributions 
deduced from optical and radio observations, respectively, do not agree. The latter 
discrepancy is discussed
in more detail in a subsequent section.

In this study, we have attempted a direct comparison by doing HI absorption measurements
towards the bright stars themselves. Of course, this can
be done only in those lines of sight (towards stars studied previously) where there
are strong enough radio sources. We were actually able to do this in about 25 directions.
From this data we obtain the velocities of the
absorbing gas, and in combination with previous emission measurements in the
same directions, the spin temperature of the HI gas in the very entities seen
in optical absorption. This is the first attempt at such a direct comparison.

Identifying the optical absorption lines with 21 cm absorption features is particularly
important because of a long standing puzzle.
Optical absorption along many lines of sight reveal two sets of absorption
features. One set occur at near zero (low) velocities and another at higher
velocities with respect to the local standard of rest. The faster
clouds have measured velocities well in excess of the radial
component of Galactic rotation one could attribute to them. 
In a classic study of Adams' data, Blaauw (1952) clearly showed the existence
of a high velocity tail extending upto as high as 100 \kms in the distribution 
of random velocities of the optical
absorption features. Thus the existence of a high velocity population of clouds
was firmly established. 
There was also a hint that these ``fast''
clouds may belong to a different population. 
They exhibited the
well known Routly-Spitzer effect (Routly and Spitzer 1952) \ie the
NaI to CaII ratios in these clouds were significantly lower than those
in the lower velocity clouds.

In the decade following the discovery of the 21 cm line attempts were
made to detect the gas seen in optical absorption.
These involved measuring HI emission
in the direction of stars which show optical absorption features in their
spectra in order to compare the HI spectra with the CaII and NaI spectra
(Takakubo 1967;  Habing 1968, 1969;
Goldstein and MacDonald 1969; Goniadzki 1972; Heiles 1974 and others).
Habing's study in
particular targeted
selected stars to attempt a direct face off. The results of these early
studies threw up another intriguing fact. The low velocity features appeared
to be well correlated in HI emission and optical absorption, \ie whenever
the optical spectra showed a low velocity absorption feature (v$<$ 10 \kms),
there was HI emission
at the corresponding velocity. However, the high velocity features were in general
{\em absent} in HI emission down to low limits (T$_b <$ 1 K). This
seemed to be the case in all directions in the sky.

The typical beam sizes in the early experiments to measure HI emission were $\sim$ 0.5$^\circ$.
The angular size of the absorbing gas could have been much smaller leading to
considerable beam dilution. This could be the reason why the clouds were not detected in HI emission.
This exemplifies the difficulties in
comparing the features seen in optical absorption with arc-second
resolution with those seen in radio emission using a comparatively
large beam. This is one of the major factors which led us to attempt
an HI {\it absorption study}. The resolution achieved is of the same order
as in optical absorption enabling a surer comparison. 

Despite the failure of the early emission measurements to detect the
``fast'' clouds there is some indication for a population of high velocity clouds
from an independent HI absorption study, though this is far from being 
firmly established. 
Radhakrishnan and Srinivasan (1980) from a detailed
analysis of the 21 cm absorption profile towards the Galactic centre
suggested that the peculiar velocities of HI clouds cannot be understood
in terms of a single Gaussian distribution with a dispersion of about 5 $-$
7 \kms, the velocity distribution for ``standard'' HI clouds.  
There was an indication of a second population of weakly
absorbing clouds with a velocity dispersion  $\sim 35$ \kms.
But this has not been confirmed by other studies. Our study is an
attempt to address this important but neglected problem.
\section{Scope of the present observations}
We chose 25 stars towards which both low and high
velocity absorption features have been seen (from CaII or NaI
atoms or both). Two of the stars HD 14134 and 14143 happen to be in the same field 
(in our radio observations) and hence we have
measured absorption in only 24 fields. The positions of all the selected stars 
as projected on the plane
of the Galaxy are shown in Fig 1. The Galactic spiral arms shown in this
\begin{figure*}
\begin{center}
\mbox{
\epsfig{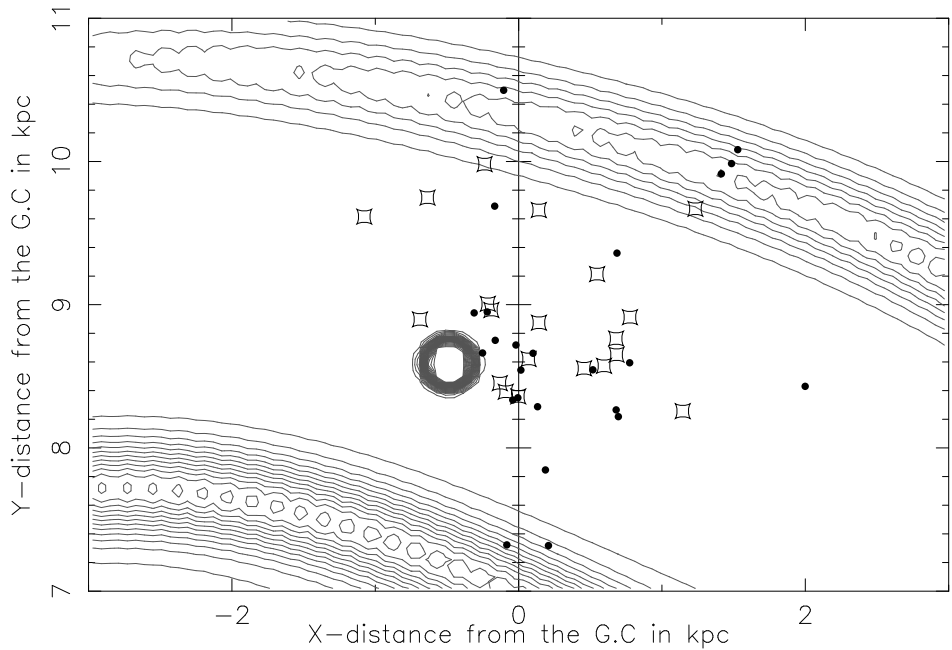}}
\end{center}
\caption[Positions of our sample of stars, projected on the Galactic plane]{24 of our
sample of
25 stars (HD 119608 lies outside the region shown) are shown
as dots on the contour plot of the
Galactic spiral arms from Taylor and Cordes. The squares show the local OB
associations from Blaauw (1985). The Sun is at 8.5 kpc from the Galactic center.
The object with dense contours to the left of the sun is the Gum Nebula.
The outer arm shown is the Perseus
arm, while the inner one is the Sagittarius arm. The three stars bunched together well
inside the Perseus arm are HD 14134, 14143 and 14818. The Perseus OB1 cluster is seen close
to them. The two stars seen projected on the Sagittarius arm are HD 166937 and 159176. 
These fields show
high velocity HI absorption.}
\end{figure*}
figure are from the electron density model by Taylor and Cordes (1993). In 13 of 
these directions
HI emission measurements had been
carried out earlier by Habing (1968,1969). Most of the fields studied by Habing were 
chosen at $|\rm b|$ $>$ 20 $^{\circ}$ to avoid complicated HI emission arising in the plane.
Four stars, however, have $|\rm b|$ $<$ 10$^{\circ}$ and were selected since they
show CaII velocities well outside the limits of Galactic rotation
at the estimated distances. Out of the remaining 12 fields, 3 have been
investigated by Goniadzki (1972) for HI emission and one by Takakubo (1967).
The rest of the fields have not been investigated in the radio.
However, we have been able to get the brightness temperatures and
column densities of HI in these directions also from the Leiden-Green Bank
survey (Burton 1985).

In the observations described in this paper
we have obtained HI absorption spectra towards a radio source whose line
of sight is close to that of the
star in each of the 24 fields.
We have been able
to identify fairly strong radio sources (in most cases $\gsim$ 50 mJy at 21 cm) within
30' to 40' of the star in every field. In more than
half the fields the radio sources are within 10' and several within 5' of the star 
in question.  If the cloud is halfway to the star, then given a 10' separation between
the star and the radio source one will sample a cloud of size less than 2 pc even
for the farthest stars in the sample which are
about 2 kpc away. This is well within  the range of accepted sizes for standard
clouds (Spitzer 1978).
In some fields we
had to arrive at a compromise between the minimum cloud size we could sample
(\ie the proximity of the projected position of the background source to the star)
and the optical depth sensitivity we could achieve. The latter constrains
us to fairly strong radio sources for reasonable integration times.
This flux density requirement limits the number of sources that one can find
close to the star and hence the trade off. Most of the background sources 
selected are from the NRAO/VLA Sky Survey (Condon et al.  1996). We aimed at 
an optical depth
sensitivity of $\tau$ = 0.1 and have done better than that in several cases.
The worst case limit on the optical depth is 0.4.

\section{Observations and data analysis}

We obtained the absorption spectra using the Very Large Array (VLA\footnote{The VLA
is operated by the National Radio Astronomy Observatory (NRAO). The NRAO is a facility
of the National Science Foundation operated under cooperative agreement by Associated
Universities, Inc.})
in the
$\sim$ 1 km and $\sim$ 11 km configurations
with a synthesized beam size of 44'' and 4'', respectively, at 21 cm.
The observations were carried out
with a total bandwidth of 1.56 MHz using both polarizations
and 128 channels.
We used 0134+329 as the primary flux calibrator. For each
source we observed a nearby secondary calibrator to do a phase and
bandpass calibration. The calibrator was observed with the frequency
band shifted by 1.5 MHz, corresponding to a velocity shift $\sim$
300 km s$^{-1}$. This shift in velocity is sufficient to move the band away
from any Galactic feature which might affect the bandpass calibration.
Typically, each of these calibrators were observed for $\sim$ 10 minutes.
The typical strengths of these calibrator sources being $>$ 1 Jy, 10 minutes of
integration time was sufficient to achieve a signal-to-noise ratio
greater than that on the source by a factor of 2.
After on line Hanning smoothing over 2 channels, the frequency resolution
obtained was $\sim$ 12 kHz
which corresponds to a velocity resolution of $\sim$ 2.5 \kms.
The integration time on
each source was chosen to give an optical depth sensitivity
of \mbox{$\tau \sim$ 0.1}, and ranged from a few minutes to more than an hour.
In all a total of \mbox{$\sim$ 30} hours were spent on the sources, split over
several sessions of observing.

The analysis was carried out using the Astronomical Image Processing System 
(AIPS) developed by the 
National Radio Astronomy Observatory (NRAO). The first step
was to make continuum images of each field and examine them for bright
sources, including the target source.
Our observations coincided with the 
ongoing NRAO/VLA Sky Survey (in the B and D configurations) and
most of our observations were carried out during the daytime. Hence we had
to contend with moderate to strong levels of interference over most of the band
owing to which approximately 10\% of the data was lost.
The next stage involved removing the continuum level from all channels. The
task used for this was UVLIN. The continuum level to be removed 
is determined from a linear fit to the visibility levels in selected channels,
which are chosen
to be free of interference as well as any spectral features. Finally the image cubes were
made. The imaging was typically done over an area of 512 by 512 pixels. In some cases 
several cubes were synthesised for different areas to cover all the
sources of interest in the field.
The spectra towards each of the sources were analysed with the Groningen Image Processing
System (GIPSY; van der Hulst et al. 1992).

\renewcommand{\arraystretch}{.9}
\begin{table}[h]
\begin{center}
\begin{tabular}{|l|l|l|l|c|c|r|}\hline
Field & d  & l$^{{\tiny II}}$, b$^{{\tiny II}}$ & $\theta$ & V$_{lsr} $ & $\tau$ & V$
_{Gal}$\\ \hline\hline
{} & pc &   & arcmin & \kms &   & {\scriptsize km s$^{-1}$}\\ \hline
14143$^*$&  2000 &  135, $-$4&  10.0&  $-$52.8, $-$50.3, $-$46.1,&  0.09& $-$25 \\
&  &           &      &  $-$11.2, $-$8.2&      &     \\
14818&  2200&  136, $-$4&  10.0&  $-$55.8, $-$14.0, $-$3.7&  0.05&  $-$25 \\
21278&  190&  148, $-$6&  37.0&  2.8&  0.08&  $>$0 \\
21291&  1100&  141, 3&  36.0 &  $-$31.2, $-$7.0, $-$4.9& 0.07&  $-$10 \\
24912&  46&  160, $-$13&  2.2&  4.3, 6.0, $-$82.0&  0.39&  $-$5 \\
25558&  220&  185, $-$33&  26.0&  8.1&  0.10&  $<$0\\
34816&  540&  215, $-$26&  40.0&  6.0, 8.0&  0.10&  $<$0 \\
37742&  500&  206, $-$16&  15.0&  9.5&  0.03&  5\\
38666&  300&  237, $-$27&  21.5&  none&  0.02&  5 \\
41335&  300&  213, $-$13&  3.0&  1.0, 10.0&  0.04&  $>$0 \\
42087&  1200&  188, 2&  42.0, 32.0&  4.4, 6.5, 12.4&  0.13&  $-$5 \\
93521&  2000&  183, 62&  27.0&  none&  0.11&  $<$0 \\
119608&  3400&  320, 43&  16.0&  $-$5.4&  0.09&  $-$20 \\
141637&  170&  346, 21&  11.0&  0.5, $-$0.2&  0.28&  $\sim$0\\
148184&  150&  358, 21&  0.82&  3.4&  0.26&  $\sim$0\\
156110&  720&  71, 36&  $<$10.0&  2.2&  0.18&  5 \\
159176&  1180&  356, 0&  5.3&  $-$20.8, $-$74.0&  0.28&  $\sim$0 \\
166937&  1200&  10, $-$2&  5.0, 7.0&  34.6, 47.2, 35.7&  0.39&  5 \\
175754&  680&  16, $-$10&  $<$10.0&  6.8&  0.09&  10 \\
199478&  2000&  88, 1&  8.0&  5.0 to $-$75.0&  0.21&  $-$7 \\
205637&  250&  32, $-$45&  18.4&  none&  0.30& 5 \\
212978&  520&  95, $-$15&  12.0&  0.3, $-$12.3&  0.04&  $-$5 \\
214680&  780&  97, $-$17&  12.0&  $-$4.8, 1.4&  0.28&  $-$5 \\
220172&  750&  68, $-$63&  2.9&  none&  0.33&  10 \\ \hline
\end{tabular}
\caption[Summary of HI absorption.]
{Summary of HI absorption: Column 1 gives the HD number of the
star. Columns 2 and 3 show the distance to the star (as listed in the Sky Catalogue 
2000.0, Vol 1, Hirshfield and Sinnott 1982) and its galactic coordinates, respectively.
Column 4 gives the angular separation of the star from the radio
source(s) towards
which absorption has been detected. Column 5 lists the
HI absorption velocities (LSR).
Columns 6 shows
the detection limit in $\tau$ (towards the
strongest source in the field).
Column 7 shows the approximate radial component of the Galactic rotation velocity at the
distance to the star.  $<$ 0 and $>$ 0 are used to indicate velocities with magnitude less 
than 5 \kms.\\
{\footnotesize The stars 14134 and 14143 are in the {\em same} field referred to above
as 14143$^*$.}}
\end{center}
\end{table}
\renewcommand{\arraystretch}{1}
\renewcommand{\arraystretch}{.9}

\begin{table}
\begin{center}
\begin{tabular}{|l|l|l|l|c|}\hline
\setcounter{table}{1}
Field & V$_{lsr}$(optical, all) & V$_{lsr}$(HI, coincident)& V$_{Gal}$&Ref \\ \hline\hline
{} & \kms  & \kms & \kms &   \\ \hline
14143$^*$&  $-$62.3, $-$46.8, $-$6.3&  $-$46.1, $-$50.3, $-$11.2& $-$25 & M \\
& $-$65.3, $-$50.8, $-$10.3& & & \\
14818& $-$42.6, $-$33.6, $-$6.6& $-$3.7& $-$25 & M \\
21278&  $-$0.2, 48.6&  2.8& $>$0 & A  \\
21291&  $-$34.0, $-$7.5&  $-$31.2, $-$7.0, $-$4.9& $-$10 & M \\
24912&  4.7, 20.7&  4.3& $-$5& A\\
25558&  10.1, 19.0&  8.1& $<$0 &  A \\
34816&  $-$14.0, 4.14&  6.0& $<$0 & A \\
37742&  $-$21.0, 3.6&  none & 5 & A \\
38666&  1.0, 20.2&  no absorption& 5 &   MZ \\
41335&  $-$20.8, 0.2&  1.0& $>$0 &  A \\
42087&  $-$37.7, $-$4.8, 10.2&  12.4& $-$5 & A \\
93521&  $-$55.0, $-$34.0, $-$10.3, 6.8&  no absorption& $<$ 0 & MZ  \\ 
119608&  1.3, 22.4&  none & $-$20 & MZ \\
141637&  $-$22.0, 0.0&  0.5, $-$0.2& 0 & B \\
148184&  14.2, 2.2&  3.4& 0 &  A \\
156110&  $-$19.7, 0.4&  2.2& 5 &  MZ \\
159176&  3.5, $-$22.5&  $-$20.8& 0 & A \\
166937&  $-$5.5, 5.9, 25.3, 41.1&  5.4, 47.2, 35.7& 5 & A \\
175754&  $-$73.0, 5.9, 29.5&  6.8& 10 & A \\
199478&  $-$2.1, 8.7, 42.3, 61.2&  3.8 (blend) & $-$7 & A  \\
205637&  $-$13.9, 1.8&  no absorption& 5 & A   \\
212978&  $-$73.0, 0.6&  0.3& $-$5 &  A \\
214680&  $-$23.7, $-$14.7, 0.1&  1.4& $-$5 & A \\
220172&  $-$21.5, $-$0.8, 13.5&  no absorption& 10 & MZ  \\\hline
\end{tabular}
\caption[Summary of coincident velocities.]{Summary of coincident velocities: Column
2 lists the LSR velocities of all the optical absorption features seen in each
line of sight (most of the listed velocities are
from CaII observations). Velocities of the matching HI absorption features are given in 
column 3.
Column 4 shows the radial component of the Galactic rotation velocity at the distance
to the star.
Column 5 has the
reference for the optical absorption velocities
which are from A: Adams (1949), B: Buscombe and Kennedy (1962), M:  M\"{u}nch (1957) and
MZ: M\"{u}nch and Zirrin (1961).
}
\end{center}
\end{table}
\section{Results}
Table 1 lists the details of all the fields observed.
In most of these fields
there were several radio sources within the primary beam in addition to the
source initially targeted. We have obtained spectra towards these
sources as well if they turned out to be strong enough to detect reasonable
optical depths ($\sim$ 0.1).
Column 1 gives the HD number of the
star. Columns 2 and 3 give the distance to the star (as listed in the SKY 2000 catalogue
of bright stars) and its galactic coordinates, respectively.
Column 4 gives the angular separation of the star from the radio
source(s) towards
which absorption has been detected. Column 5 lists the
HI absorption velocities (LSR).These velocities have been derived by fitting Gaussian profiles
to the absorption features.
Columns 6 shows
the detection limit in $\tau$ (towards the
strongest source in the field).
Column 7 shows the radial component of the Galactic rotation velocity at the
distance to the star.

The optical absorption
velocities are likely to suffer from blending
of features due to lack of resolution. A discussion of this and associated
problems can be found in Welty, Morton and Hobbs (1996). Moreover, the
correction for the solar motion adopted by different authors may
differ leading to errors of $\sim$ 1 \kms (see for example Blaauw, 1952). 
In some cases, we have smoothed
the absorption spectra to a resolution of 5 \kms to facilitate the convergence of the
gaussian fit. Narrow absorption features
are known to have widths less than this (Crovisier 1981). We have checked
the unsmoothed data to ensure that there are no serious effects of blending
in the estimates for velocities for the features discussed below. However,
crowding in velocities inevitably causes some of the absorption widths
to be suspect. The formal error in the fitting procedure for our HI absorption
velocities is in most cases $\sim$ 1 \kms or less.  The errors
add to \lsim 2 \kms.
However, it must be stated that the blending of features in both optical and radio
could easily lead to larger errors than this formal value. {\it Hence we consider
a feature in the optical spectrum as ``coincident'' with one seen
in HI absorption or emission if the magnitude of the difference in velocities 
is less than 3 \kms}.
This is roughly the
same criterion adopted by Habing (1969) and Howard et al. (1963) for comparing
HI emission profiles with optical absorption.

Table 2 lists the ``matching'' features \ie HI absorption features whose velocities
agree with the velocity of the optical absorption feature.
Column 2 shows the LSR velocities of all the optical absorption features seen in each
line of sight (most of the listed velocities are
from CaII observations). Velocities of the matching HI absorption features follow in column 3.
Column 4 shows the radial component of the Galactic rotation velocity at the distance
to the star to facilitate comparison with the optical absorption velocities.
Column 5 has the
reference for the optical absorption velocities
which are from Adams (1949), M\"{u}nch (1957) and
M\"{u}nch and Zirrin (1961). We have used later compilations of these
by Takakubo (1967), Siluk and Silk (1974), and Habing (1969).

We present the HI optical depth towards all the observed fields
at the end of the paper.
The coordinates of the radio source (epoch 1950) towards which the spectrum is
obtained is labelled at the top right of each panel. The HD number of the star 
towards which the corresponding optical spectrum is obtained is at the top left.
The star co-ordinates (epoch 1950) are given immediately below this. The star 
co-ordinates have not been repeated if there are several radio sources in the same field.
In all the spectra the velocities at which optical absorption is seen is indicated
by arrows on the velocity axis (top).
The conclusions we draw from these observations and a detailed discussion of the
implications for the models of the ISM are presented
in paper II.
\section*{Acknowledgement}
We wish to thank Adriaan Blaauw for extensive discussions which led to this
investigation. His continued interest and critical comments have been
invaluable to us. We are also indebted to Hugo van Woerden for guiding us
through the literature pertaining to the early HI observations by the Dutch group.

\begin{figure*}
\vspace{1cm}
\epsfig{file=fig_0.ps,width=12.5cm}
HI optical depth $\tau$ versus V$_{\rm LSR}$. The arrows indicate the
velocities of the optical absorption lines towards the star indicated
above each panel.
\end{figure*}
\begin{figure*}
\vspace{2cm}
\epsfig{file=fig_4.ps,width=12.5cm}
\end{figure*}
\begin{figure*}
\vspace{2cm}
\epsfig{file=fig_8.ps,width=12.5cm}
\end{figure*}
\begin{figure*}
\vspace{2cm}
\epsfig{file=fig_12.ps,width=12.5cm}
\end{figure*}
\begin{figure*}
\vspace{2cm}
\epsfig{file=fig_16.ps,width=12.5cm}
\end{figure*}
\begin{figure*}
\vspace{2cm}
\epsfig{file=fig_20.ps,width=12.5cm}
\end{figure*}
\begin{figure*}
\vspace{2cm}
\epsfig{file=fig_24.ps,width=12.5cm}
\end{figure*}
\begin{figure*}
\vspace{2cm}
\epsfig{file=fig_28.ps,width=12.5cm}
\end{figure*}
\begin{figure*}
\vspace{2cm}
\epsfig{file=fig_32.ps,width=12.5cm}
\end{figure*}
\begin{figure*}
\vspace{2cm}
\epsfig{file=fig_36.ps,width=12.5cm}
\end{figure*}
\begin{figure*}
\vspace{2cm}
\epsfig{file=fig_40.ps,width=12.5cm}
\label{lastpage}
\end{figure*}
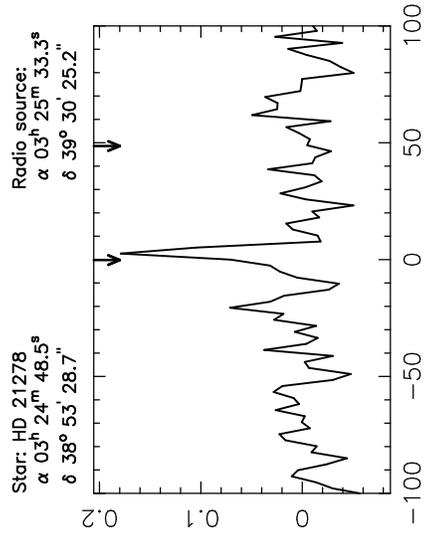
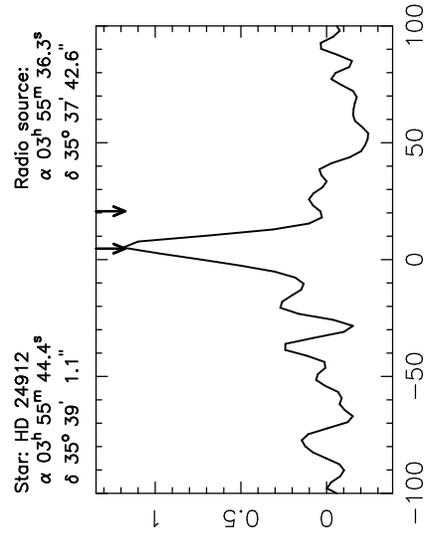
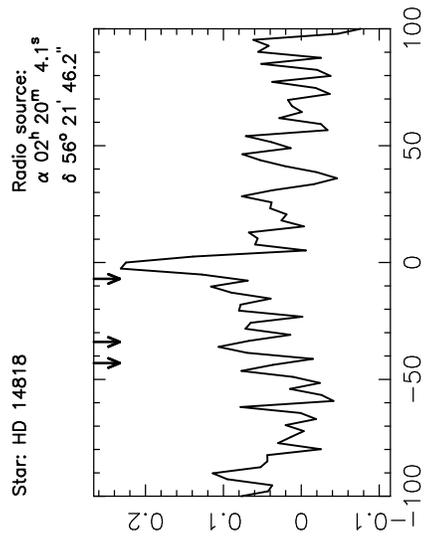
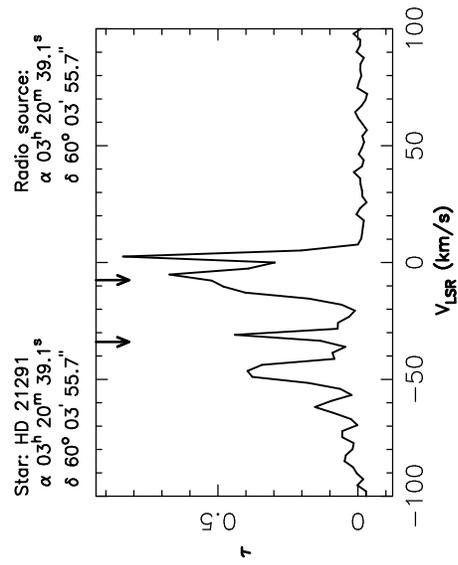
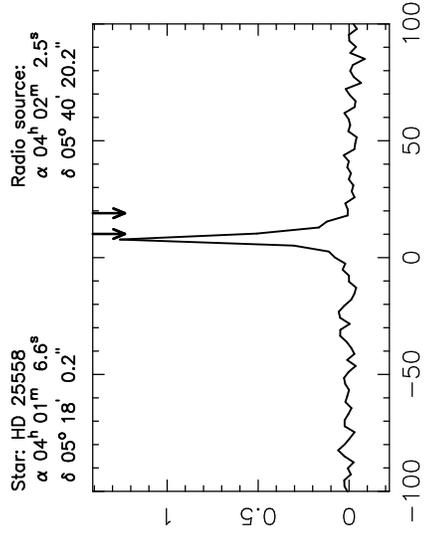
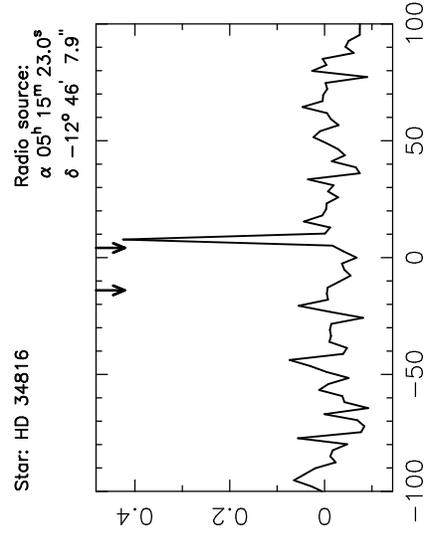
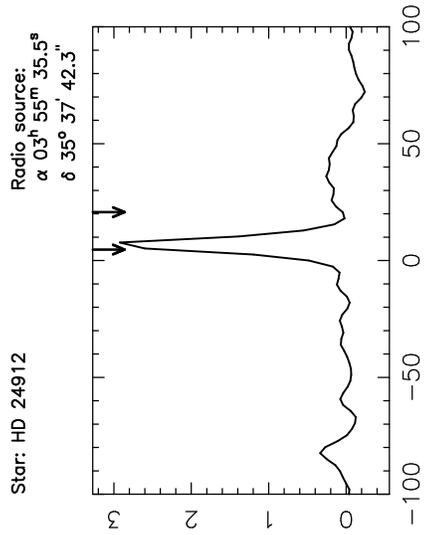
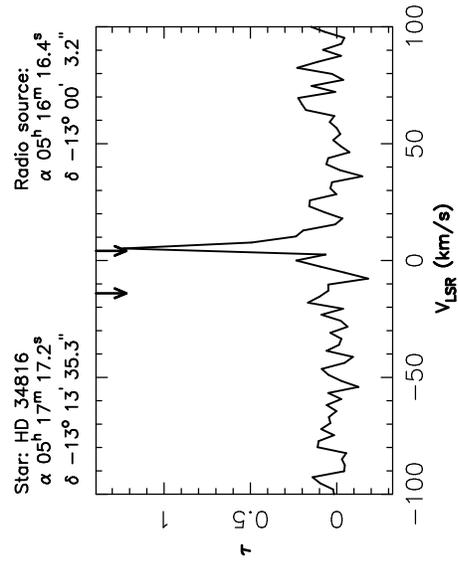
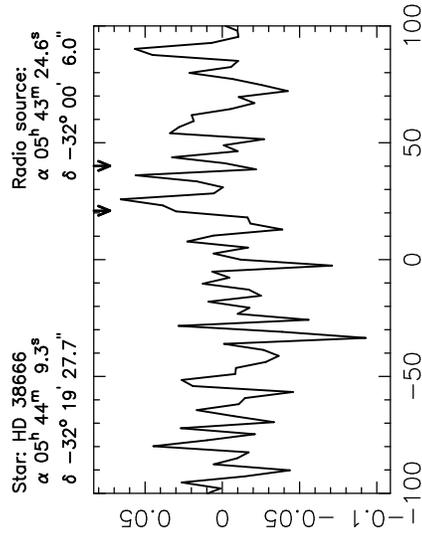
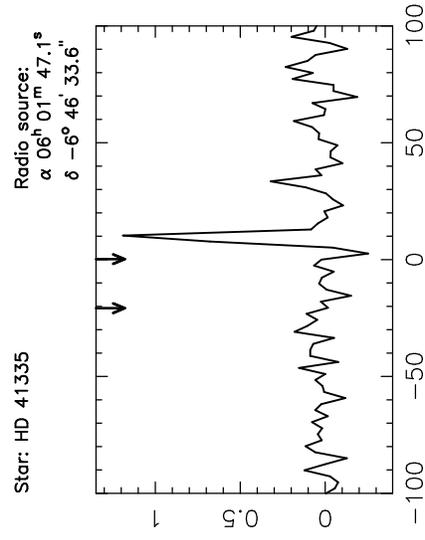
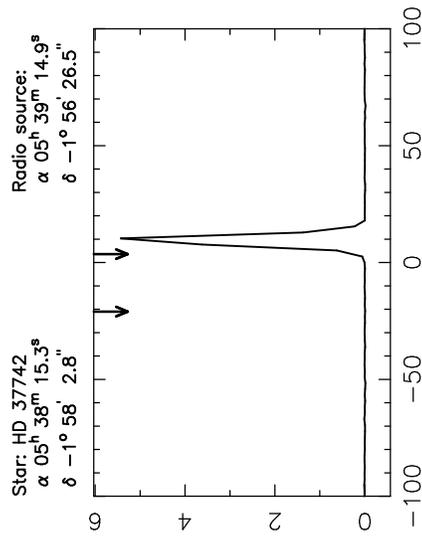
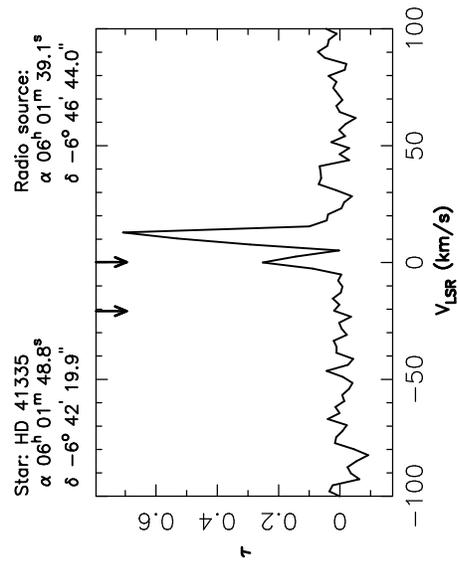
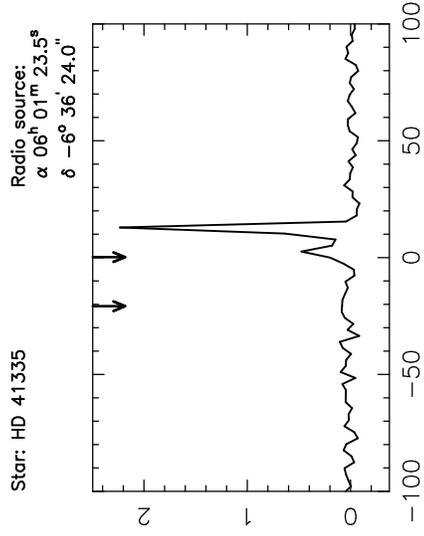
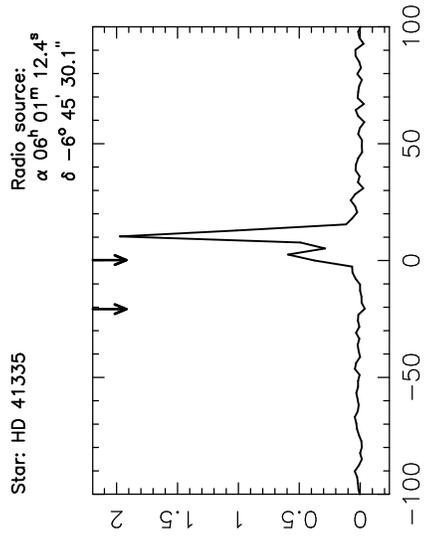
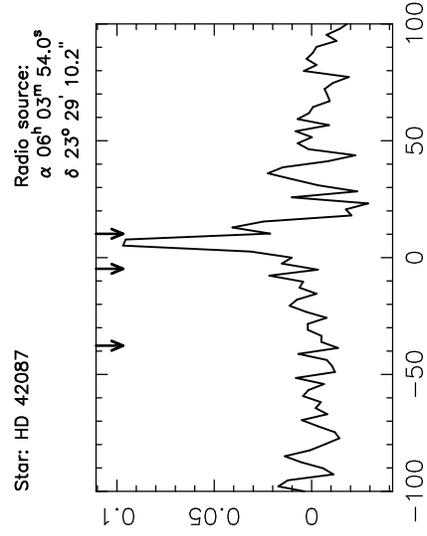
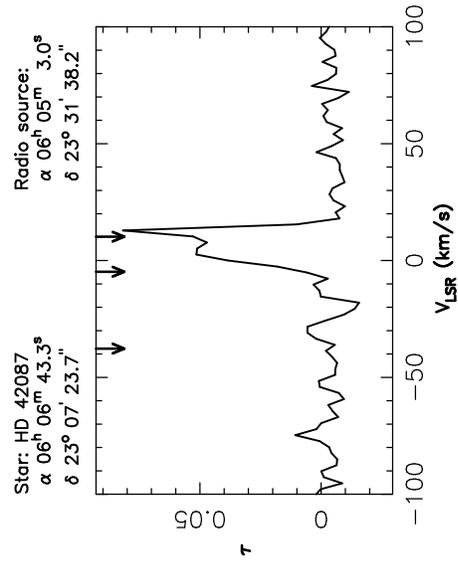
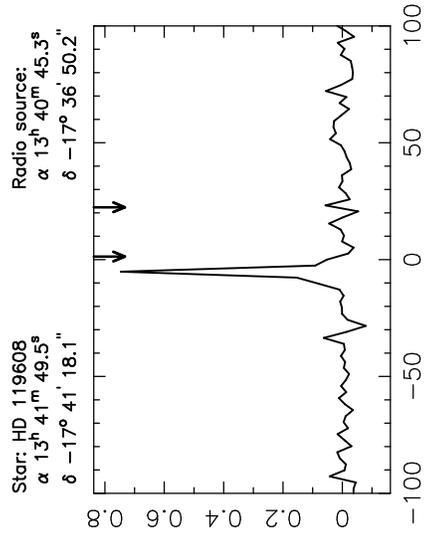
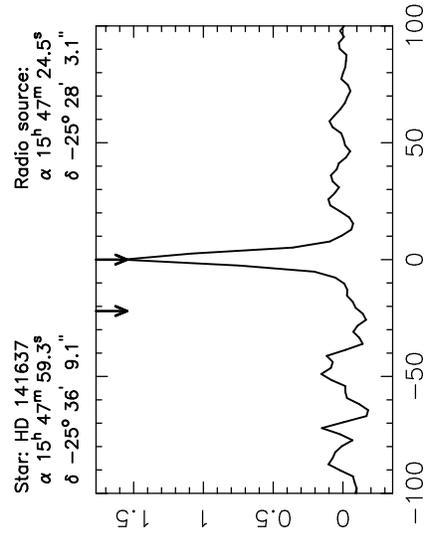
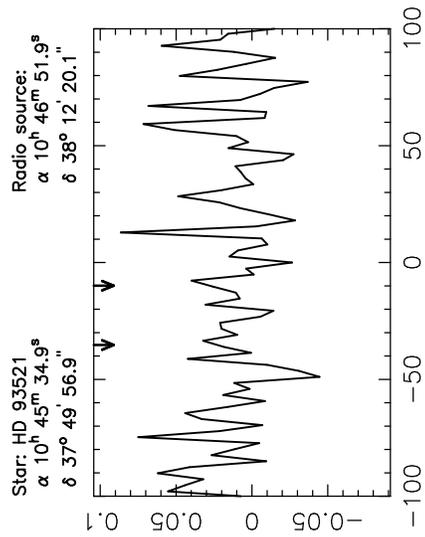
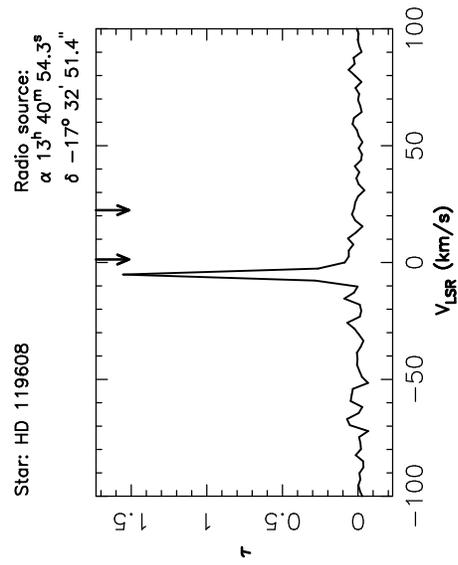
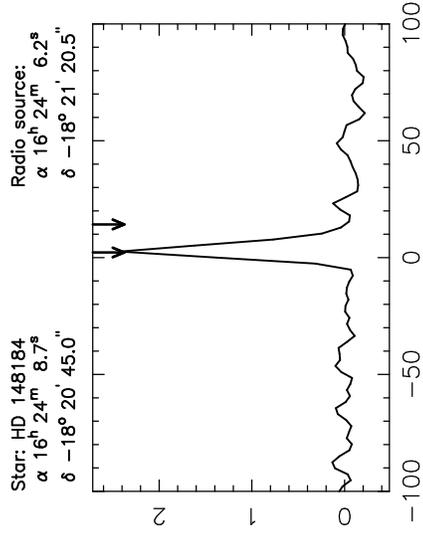
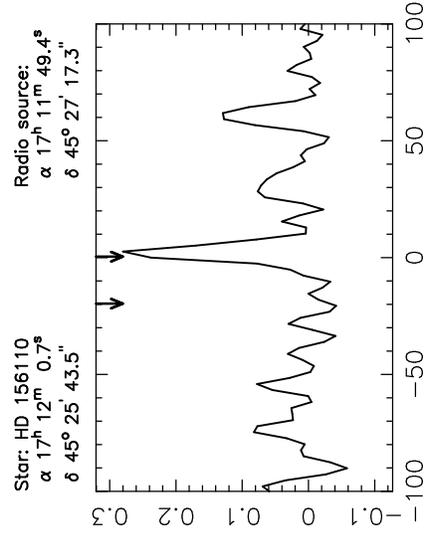
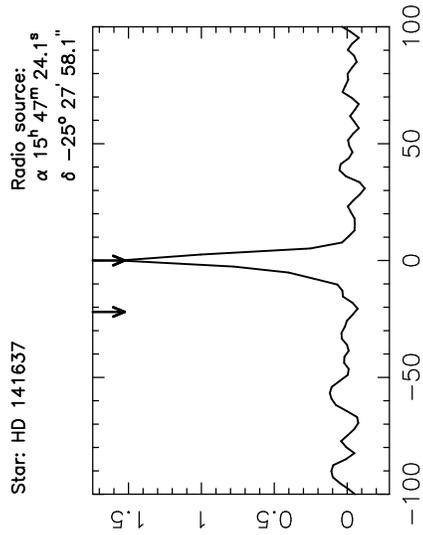
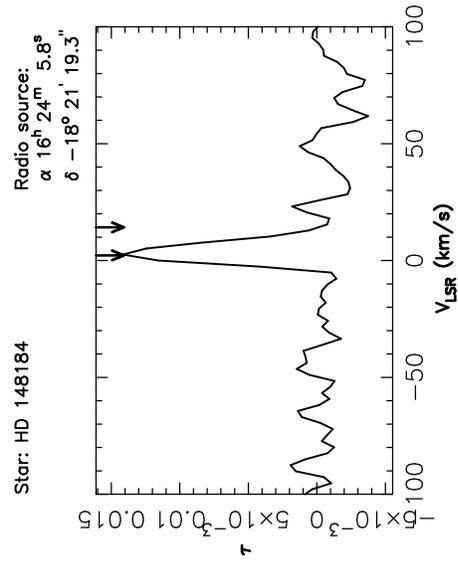
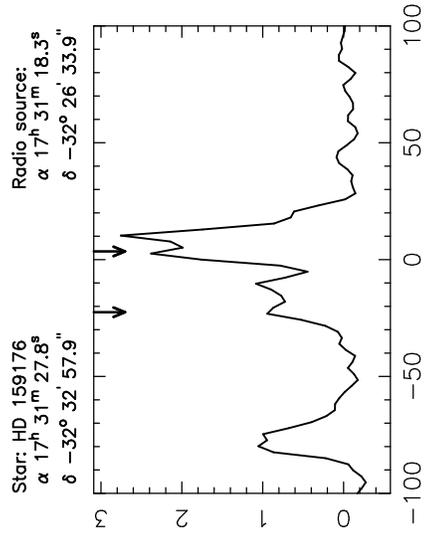
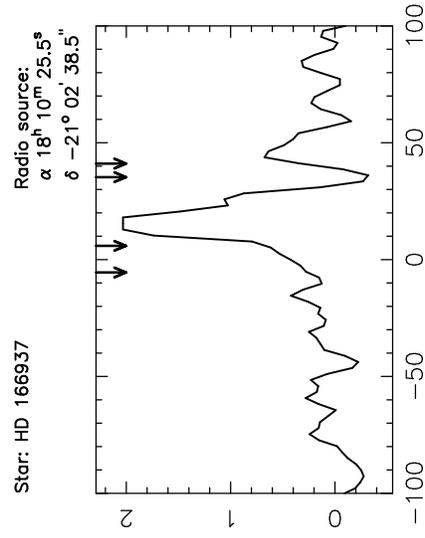
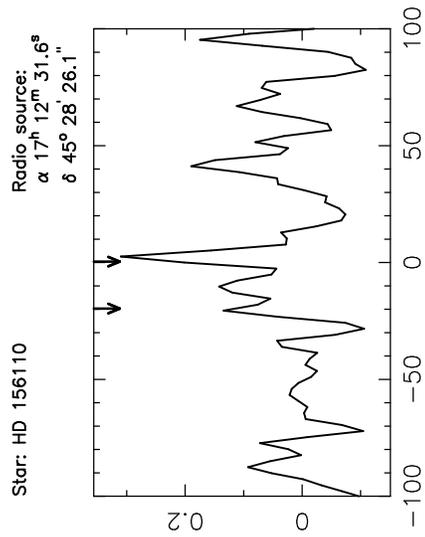
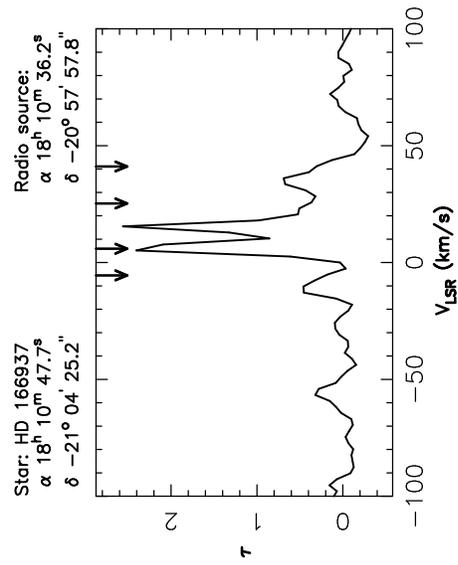
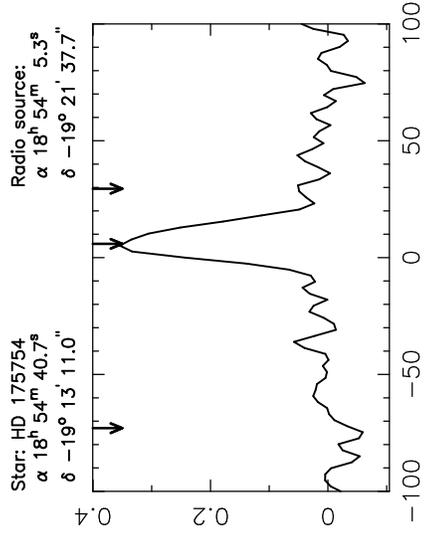
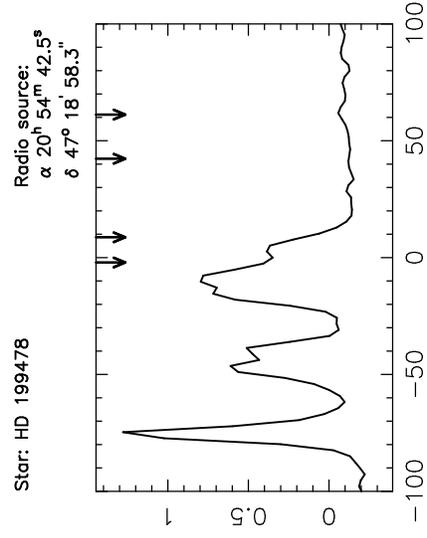
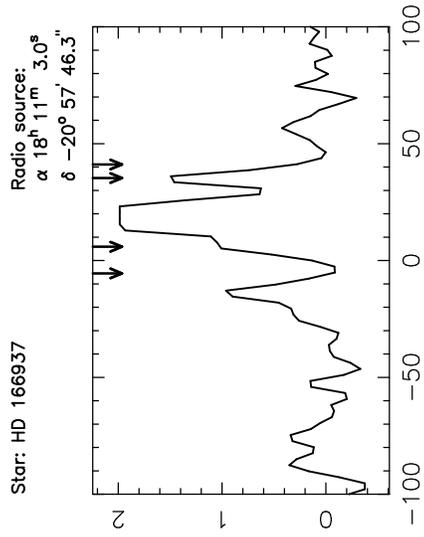
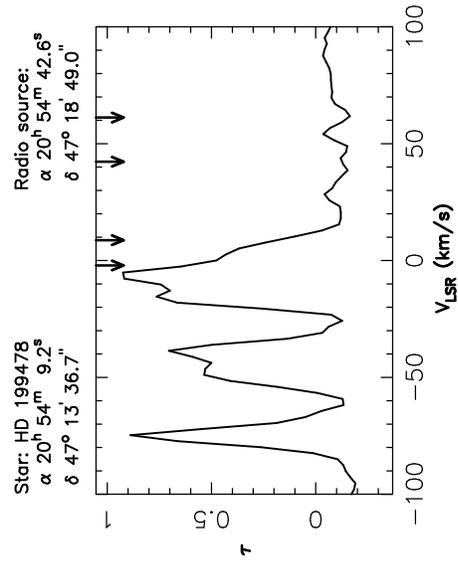
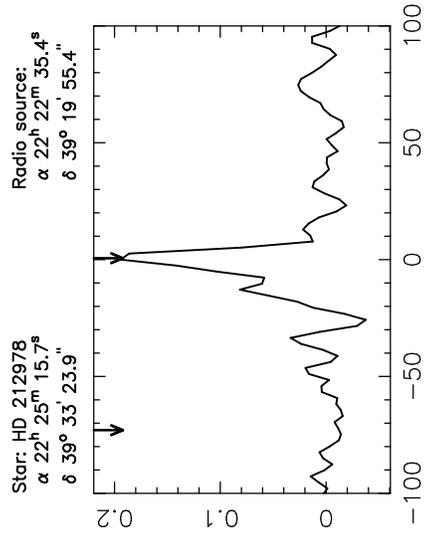
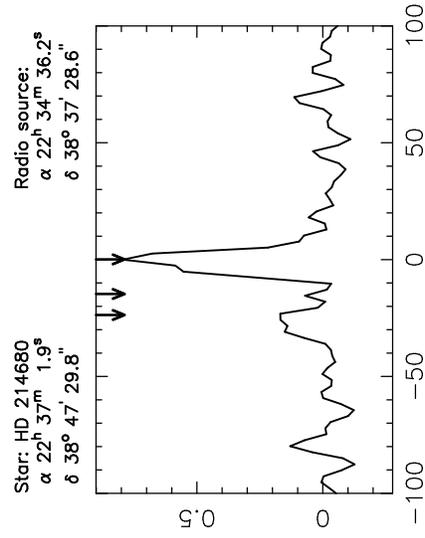
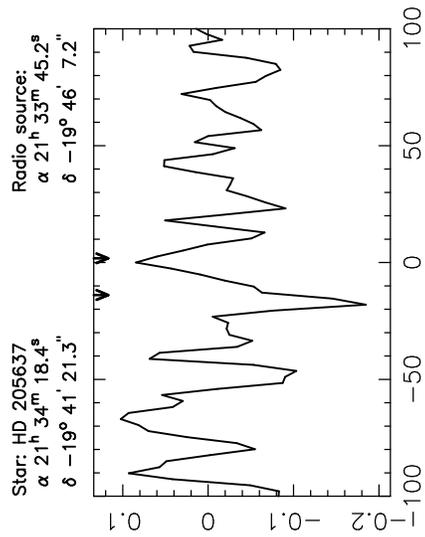
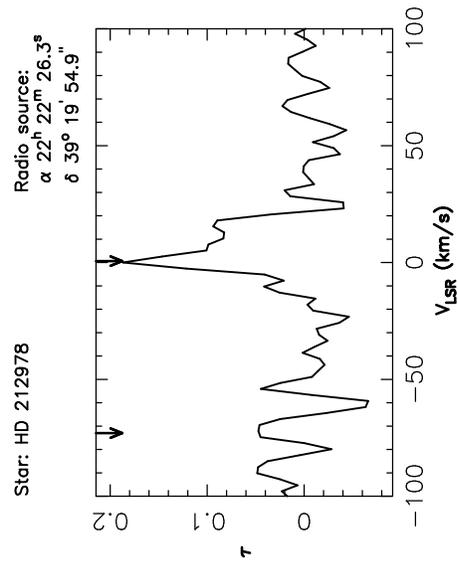
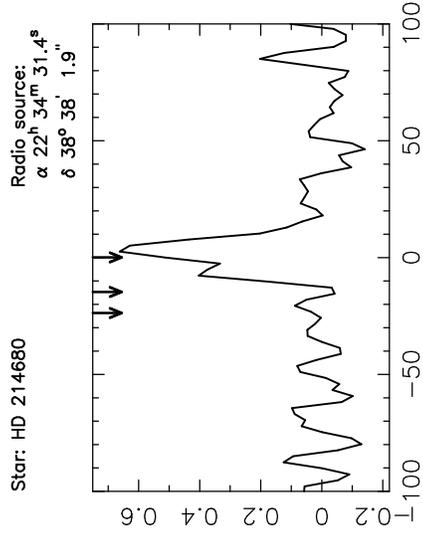
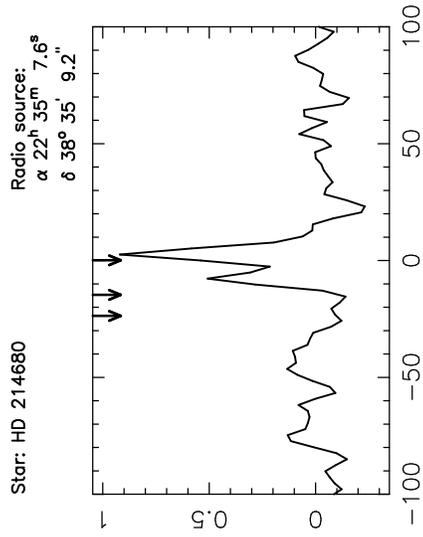
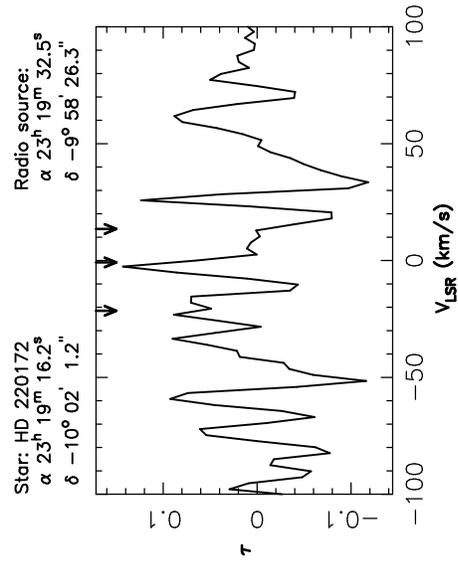
\end{document}